\documentstyle[12pt,aaspp4]{article}
\begin{document}

\title{Experimental f-value and isotopic structure for the Ni I line blended with [OI] 
at 6300\AA}

\author
{ S. Johansson \altaffilmark{1}, U. Litz\'{e}n\altaffilmark{1}, H. Lundberg 
   \altaffilmark{2},  Z. Zhang \altaffilmark{2}}


\altaffiltext{1}{Atomic Astrophysics, Lund Observatory, Lund University, 
            PO Box 43, SE-22100 Lund, Sweden; 
            email: sveneric.johansson@astro.lu.se, ulf.litzen@astro.lu.se  }
\altaffiltext{2}{Department of Physics, Lund Institute of Technology, 
            PO Box 118, SE-22100 Lund, Sweden;
            email: Hans.Lundberg@fysik.lth.se}

\date{Received <date> / Accepted <date>}


\baselineskip 0.33in
\parindent 0.25in  
\begin{abstract}{We have measured the oscillator strength of the Ni I line at 6300.34 
\AA, which is known to be blended with the forbidden [O I] $\lambda$6300 line, used for
determination of the oxygen abundance in cool stars. We give also wavelengths 
of the two isotopic line components of $^{58}$Ni and $^{60}$Ni derived from the  
asymmetric laboratory line profile. These two line components of Ni I have to be 
considered when calculating a line profile of the 6300 \AA\ feature 
observed in stellar and solar spectra. We also discuss the labelling of 
the energy levels involved in the Ni I line, as level mixing makes the theoretical 
predictions uncertain. }
\end{abstract}


\section{Introduction}

Allende Prieto, Lambert \& Asplund et al (2001, hereafter APLA) have
recently reviewed the various ways to determine the oxygen abundance in the
sun and other late-type stars and discussed the different problems connected
with the spectral oxygen lines used. Stellar conditions limit the number of
appropriate lines to a few allowed and forbidden lines of neutral oxygen (O
I and [O I]) in the optical wavelength region and molecular OH in the UV and
IR regions. A frequently used method to determine the oxygen abundance
in cool stars and the sun is to use the forbidden [O I] line at 6300 \AA ,
which is particularly studied and analysed in the solar spectrum by APLA.
However, this line is associated with problems as it is blended with an
Ni I in the solar spectrum. This was early pointed out by Lambert (1978).
Since the [OI] line (6300.31 \AA ) and the blending Ni I line (6300.34 \AA )
appear as a totally unresolved feature in the solar spectrum APLA constructed a 
spectral profile from laboratory data (wavelengths and f-values) and compared with
the observed solar feature. Three of four crucial atomic parameters are
known to a satisfactory accuracy, viz. wavelength (Eriksson 1965) and
f-value (Storey \& Zeippen 2000) for the [OI] line and wavelength for the Ni
I line (Litz\'{e}n, Brault \& Thorne 1993, hereafter LBT). APLA used a three-
dimensional time-dependent hydrodynamical model to simulate the solar surface 
and applied the same technique as previously used by Asplund et al. (2000) in 
the determination of the solar iron abundance. They used three
free parameters, the continuum level, the oxygen abundance and the product ''
$gf$-value of the Ni I line x nickel abundance, $gf\epsilon$(Ni)''
to match the predicted and observed profiles. By inserting the
adopted solar abundance of nickel from Grevesse and Sauval (1998) APLA
derived an ''astrophysical'' log $gf$ value of -2.31 for the Ni I
line from the fitted value of $gf\epsilon$(Ni). We quote from the APLA
paper: ''The log $gf$ for the Ni I line is uncertain. There are
seemingly no laboratory measurements for this line''.

We have now determined the $gf$-value of the Ni I line at 6300.34 \AA\
by combining two-step laser-induced fluorescence (LIF) measurements of the
radiative lifetime of the upper level with branching fraction measurements
using Fourier Transform Spectroscopy (FTS). We have also fitted two isotopic 
line components ($^{58}$Ni and $^{60}$Ni) to the laboratory Ni I line and derived
wavelengths and absolute intensities for both components. These two Ni isotopes
account for 94\% of the solar nickel abundance. We have reexamined the LS
composition of the upper energy level of the Ni I transition, as it is
severly mixed and has no clear LS signature. The level was discussed and
reassigned in the extensive work on the Ni I spectrum by Litz\'{e}n, Brault
\& Thorne (1993), and its identity has been further discussed by APLA.
Adopting one or another of the "old" (4d e$^{3}$P$_{0}$) or the "new" 
(4s$^{2}$ $^1$S$_{0}$) assignment makes a difference in the $f$-value of 
a factor of 400 (a difference in log $gf$ of 2.6) if we consult the
 Kurucz database, which is often used in abundance work using the spectrum 
synthesis technique. Thus, an experimental value of the oscillator strength 
will help in understanding the real LS composition of the energy level.

\section{Atomic physics background}

Since the Ni I line studied in this paper has a great influence on the
determination of the oxygen abundance it deserves a special attention. Because
of the level mixing there is a need for an experimental $gf$-value as well as 
a detailed study of the isotopic composition of the line.
The effect of level mixing that makes calculated oscillator
strengths very uncertain is a frequent problem in complex spectra. Very
drastic cases may occur in modelling stellar spectra where a calculated
spectral line, predicted by means of theoretical atomic data, totally
disagrees with the observed feature.

To illustrate the level mixing in the present Ni I case we have included a
small part of the Ni I term diagram in Fig. 1, showing the relevant energy
levels involved in the discussion as well as in the measurements. The upper
level of the $\lambda$6300 Ni I line is located at 50276 cm$^{-1}$ and the lower
level is y$^{3}$D$_{1}^o$, which belongs to the odd parity 3d$^{8}$4s4p
configuration. In the first analysis of the Ni I spectrum, Russell (1929)
assigned the 50276 level to e$^{3}P_{0}$ of the even parity 3d$^{9}$4d
configuration, which seemingly makes the transition to y$^{3}D_{1}^o$ a
''two-electron jump''. The appearance of such a transition can be explained
by configuration interaction between 3d$^{8}$4s4p and 3d$^{9}$4p, making the
line a participant of a regular 4p-4d transition. Theoretical calculations by
Litz\'{e}n et al. (1993) confirmed such a level mixing between y$^{3}$D$_{1}^o$
and z$^{3}$D$_{1}^o$ belonging to 3d$^{8}$4s4p and 3d$^{9}$4p, respectively.

\begin{figure}[h]
\plotone{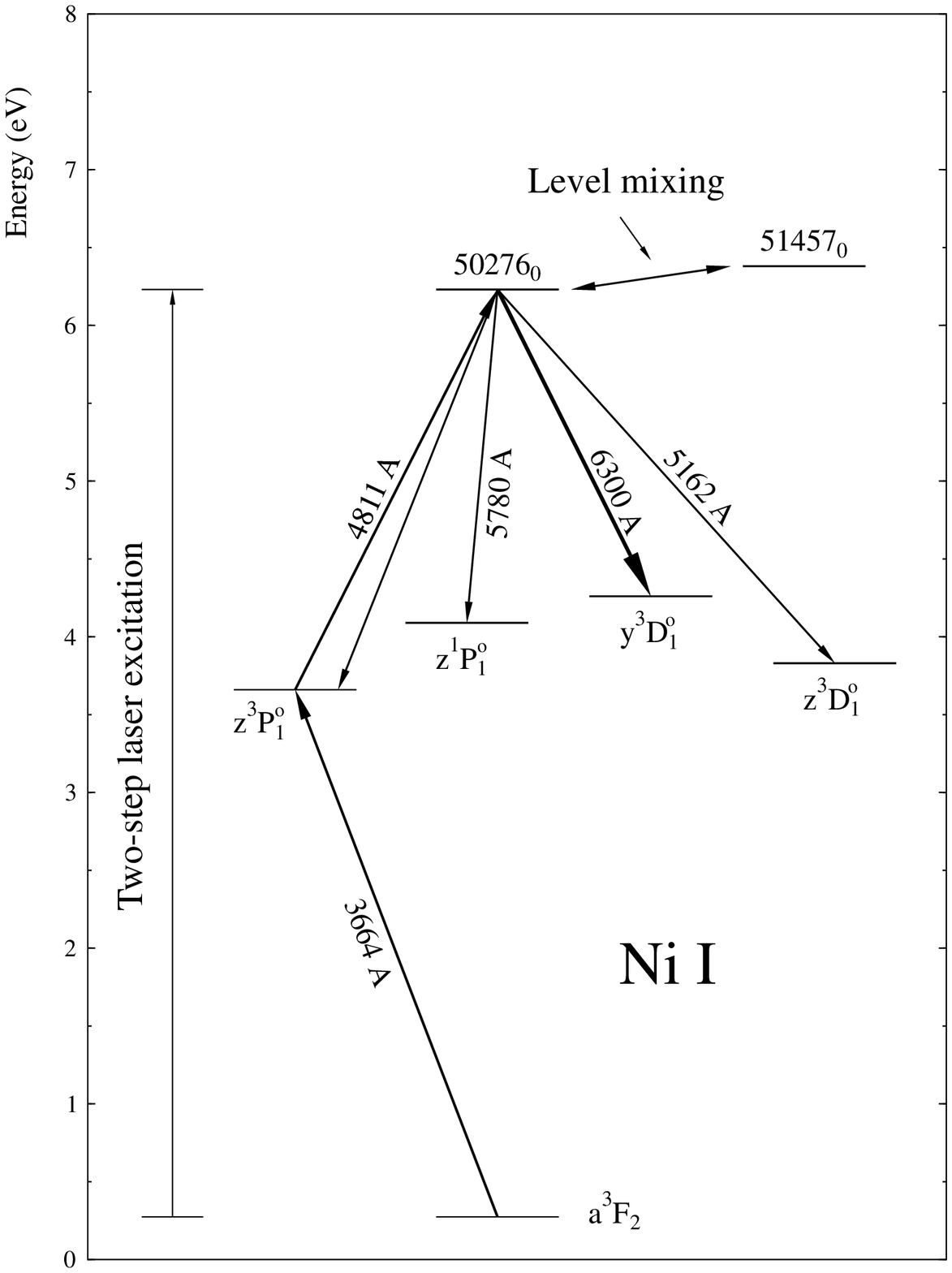}
\caption{Partial level diagram of Ni I showing the levels and transitions discussed in
this paper. The two-step excitation in the lifetime measurement is indicated to the left
} 
\end{figure}

\ In the NIST\ compilation of iron group elements (Corliss \& Sugar 1981),
issued before the LBT work, the 50276 level has been given the old label,
4d e$^{3}P_{0}$, suggested by Russell. However, the label of another level 
located at 51457 cm$^{-1}$ was changed from 4d $^{1}$S$_{0}$
to 4s$^{2}$ $^{1}$S$_{0}$. This change is supported by Kurucz's calculations
(2002) as the major eigenvector component is more than 90\% of 4s$^{2}$ $^{1}$S$_{0}$. 
The Kurucz calculation also confirms the "old" label of the 50276 level, as it was
found to contain more than 95\% e$^{3}$P$_{0}$. Strangely enough,
the Cowan code calculations performed by LBT gave completely opposite
results concerning these levels. The 50276 level contained a 90\% component 
of 4s$^{2}$ $^{1}$S$_{0}$, and the 51457 level was reassigned to the 4d
configuration. LBT give the 4d labels in jK coupling notation, which is
clearly justified by the level structure (see Fig. 3 in their paper).

According to Kurucz's calculations of log $gf$ values both levels
(50276 and 51457) have their strongest decay to z$^{3}$P$_{1}^o$ (see
Fig. 1). However, the transitions to y$^{3}$D$_{1}^o$ differ by a factor 400
in the log $gf$ values, which are -1.73 for the transition from 50276 and -4.38 
from the 51457 level. The second strongest decay from the 51457 level is to z$^{1}$P$%
_{1}^o$ according to Kurucz, but the corresponding line from 50276 should be
very weak. However, the latter is observed in the laboratory spectrum by LBT. Thus,
the 50276 level obviously has a significant singlet character, which might be
difficult to predict with sufficient accuracy in the calculations. Therefore, 
we have measured the lifetime and the log $gf$-values, and analysed the 
isotopic structure.

\section{Lifetime measurements}

In the measurements of the radiative lifetime the 50276 level was populated by 
applying a two-step pulsed laser excitation with z$^{3}$P$_{1}^o$ 
as the intermediate level according to the scheme in Fig. 1. The radiative 
lifetime was derived by time-resolved observation of the fluorescence light released 
when the 50276 level decays to the lower odd levels.
The experimental set-up was similar to the one described and illustrated in
a recent paper by Nilsson et al. (2000). Free nickel atoms were created by laser 
ablation and excited to the level investigated by laser pulses from two Nd:YAG 
laser pumped dye lasers operating in the red spectral region. The wavelength of 
the first step excitation (3664 \AA) was 
reached after frequency doubling and of the second step excitation (4811 \AA) after 
Raman shifting in hydrogen gas. The two 10 ns pulses coincided in time and space 
during the interaction with the nickel atoms. Fluorescent light was detected using a 
monochromator and a fast photomultiplier. A digital transient recorder performed 
the data acquisition. The monochromator was set on one of the decay channels shown in 
Fig. 1. Most of the recordings were taken on the strong 4811 \AA\ line. An average of 
1000 single decay events was typically necessary for obtaining a good signal-to-noise 
ratio in the exponential decay curves. Thirty curves were recorded and the result 
for the lifetime of the 50276 level is 77$\pm7$ ns. The uncertainty of the lifetime is 
a combination of statistical and, although carefully checked, possible systematic 
errors. The standard deviation of the 30 different measurements is less than 20\% of 
the uncertainty.

\section{Branching fractions and oscillator strengths}

We have recorded the Ni I spectrum with the Lund UV Fourier Transform Spectrometer (FTS)
and also extracted spectra from the Kitt Peak FTS database, previously used in the 
extensive study of the Ni spectrum by Litz\'{e}n et al. (1993). All spectra are from 
hollow cathode lamps at various running conditions (pressure and DC current). The spectra
have been intensity calibrated by means of branching ratios for internal argon lines 
(Whaling 1993), i.e. argon lines produced by the carrier gas in the nickel hollow cathode
lamp. The branching fractions have been derived from the calibrated intensities of the 
spectral lines corresponding to the four decay channels indicated in Fig. 1. 

Since the lower levels in all transitions studied have short radiative lifetimes there 
is no need for any corrections 
for self absorption in the light source. According to predictions and observations the
transitions measured account for more than 99\% of the decay from the 50276 level, and no
residual branching fraction has to be considered. The total decay rate, given by the 
inverse value of the measured lifetime, has therefore been distributed among the four
transitions in accordance with the experimental branching fractions. The results are 
given in Table 1. The uncertainties in the f-values are determined according to a 
procedure suggested by Sikstr\"{o}m et al. (2002), and they include estimated errors in
lifetime measurements, intensity measurements and the instrumental response function.

\begin{table}
\caption{Branching fractions (BF) and oscillator strengths (log $gf$)
for Ni I lines from the level at 50276 cm$^{-1}$ with a radiative lifetime of 77 ns.}
\begin{tabular}{cccccccc}
\hline\hline
Lower level & $\lambda$ & BF  & gA & \multicolumn{3}{c}{log $gf$-value}  & unc. \\ 
\cline{5-7}
&   (\AA )& & (10$^{7}$ s$^{-1}$) & this work & Kurucz & APLA & \% \\
\hline 
3d$^8$4s4p y$^{3}$D$_{1}$ & 6300.341$^a$ & 0.10 &  0.13 & -2.11 & -1.73 & -2.31 & 14 \\ 
3d$^9$4p z$^{1}$P$_{1}$ & 5780.728 & 0.10 &  0.13 & -2.18 & -3.02 &  & 14 \\ 
3d$^9$4p z$^{3}$D$_{1}$ & 5162.913 & 0.02 & 0.03 & -2.93 & -1.93 &  & 28 \\ 
3d$^9$4p z$^{3}$P$_{1}$ & 4811.983 & 0.78 & 1.01 & -1.45 & -1.48 &  & 10 \\
\hline
\end{tabular}
 $^a$Data for the isotopic line components: $\lambda$($^{58}$Ni)=6300.335(1), 
$\lambda$($^{60}$Ni)=6300.355(2); both have log $gf$=-2.11
\end{table}

\section{Discussion}

There are some conclusions that can be drawn from Table I. Firstly, the
''astrophysical'' log $gf$ value of the $\lambda$6300 line derived by ALPA 
in the fitting of the synthesized profile of the combined [OI] and Ni I 
lines to the observed
solar line is closer to the present measurement than to the value given in 
the Kurucz database. The
''astrophysical'' log $gf$-value differs by 0.20 dex ($\approx$ 60\%) from the 
new laboratory value, which has an uncertainty of about 0.05 dex (15\%). Secondly, 
the triplet content of the level 50276 is
overestimated in the calculations by Kurucz and perhaps underestimated in
the calculations in LBT. Another strong evidence for a substantial
contribution of the 4s$^{2}$ $^{1}$S$_{0}$ state to the 50276 level is the
isotope shift observed by LBT for the two strong lines $\lambda\lambda$4811,5780 
in Table I, for which the the lower level belongs to the 4p configuration. 
Using Fabry-Perot interferometry Schroeder and Mack (1961) made a detailed study
of the isotope shift in the lower configurations of Ni I. They found appreciable 
line shifts in transitions where the upper and lower configurations differ in 
the number of d-electrons. By calculating the normal mass shift and estimating 
the field shift they attributed the major part of these large shifts to specific
mass shift. The absence of a significant isotopic shift
in the 6300 \AA\ line is thus consistent with a major contribution of 
3d$^{8}$4s$^{2}$ $^{1}$S$_{0}$ to the 50276 level, as the configuration of the 
lower level, 3d$^{8}$4s4p, contains the same number of d-electrons. However, the 
observed profile of the laboratory 6300 \AA\ line shows an asymmetry that could 
be due to isotopic structure. A small isotope shift would indicate a mixing of the 
upper state with a 3d$^{9}$4d level or a mixing of the lower state with a 3d$^{9}$4p 
level or a combination of the two. We have therefore examined the line more closely.

\begin{figure}[h]
\plotone{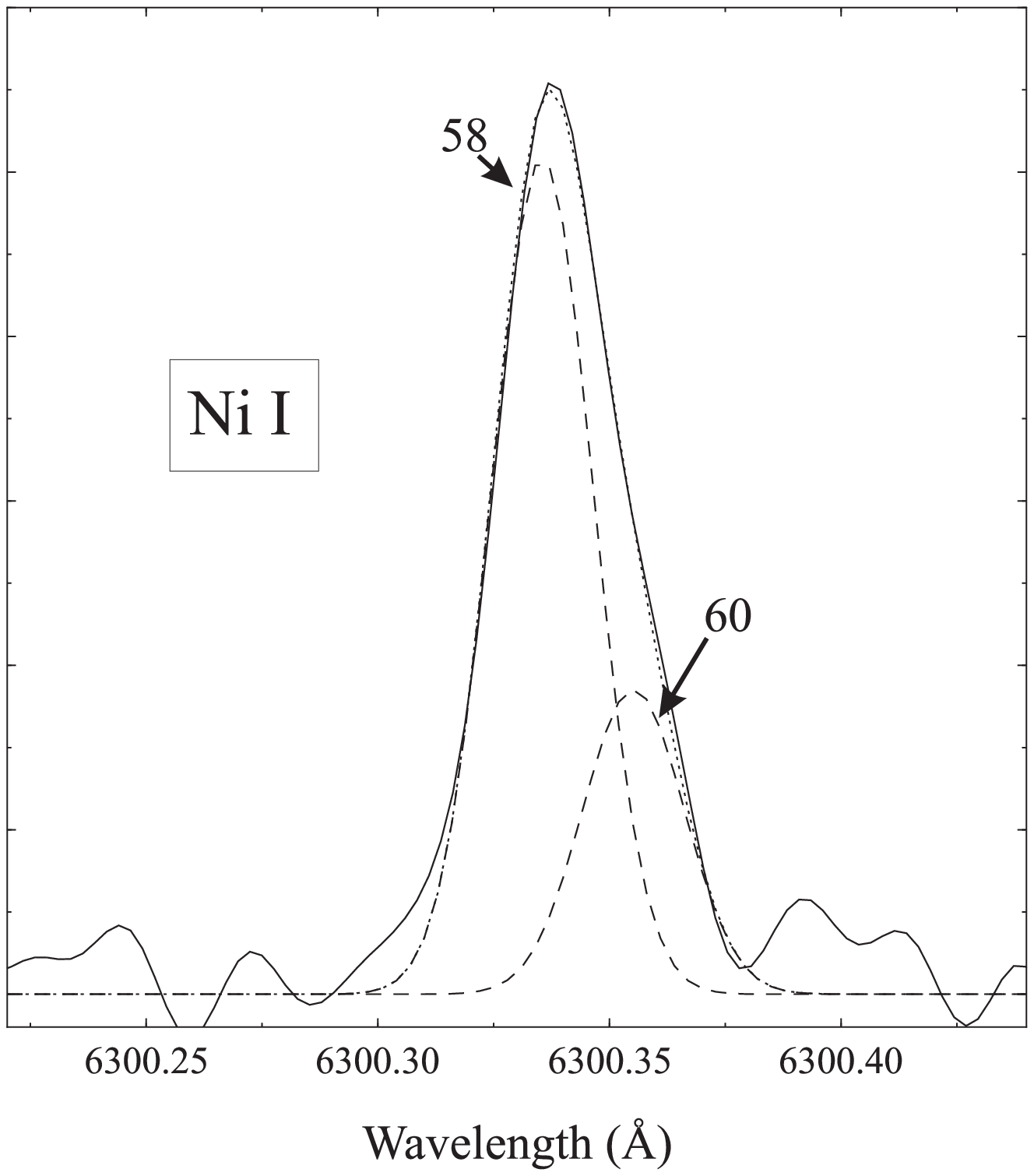}
\caption{Laboratory FTS recording of the $\lambda$6300 line of Ni I with fitted 
line components for isotopes $^{58}$Ni and $^{60}$Ni.} 
\end{figure}

In Figure 2 we show the 6300 \AA\ line as observed in the laboratory Fourier transform
spectrum. We have fitted two isotopic line components to the observed feature, by 
assuming a solar ratio of 0.38 for the abundances of the $^{60}$Ni and $^{58}$Ni 
isotopes. We have neglected the other stable isotopes $^{61}$Ni, $^{62}$Ni, and $^{64}$Ni
as they contribute only 1.1 \%, 3.6 \% and 0.9 \%, respectively. Furthermore, $^{61}$Ni 
is smeared out by hyperfine structure. The wavelengths of the two isotopic components
in the Ni I $\lambda$6300 line are 6300.335(1) for$^{58}$Ni and 6300.355(2) for$^{60}$Ni.
The uncertainties of 1 and 2 m\AA\, respectively, include errors in the fitting of the
line profile and calibration of the laboratory spectrum.  
The isotope shift of 20 m\AA\ is about the same as the distance of 25 m\AA\ between the [OI] 
line and the center of gravity (c.g.) wavelength for the Ni I 
line. Based on the present analysis we have derived new data for the c.g. of the 
line (previous values in parentheses are from LBT):
$\sigma$= 15867.769 (.773) and $\lambda$=6300.341 (.339), and inserted the new c.g. 
wavelength in Table 1. The log gf value (-2.11) is the same  for each isotopic line 
component.

The choice of LS label on the 50276 level is not evident, but it is quite
clear that the 51457 level belongs to the 4d configuration. This level does
not show any isotope shift in the transitions to 4p, which is characteristic
for all transitions measured by LBT where there is no change in the number of 
d-electrons. Considering the apparent labeling of the 51457 level and the observed 
isotope shift in the lines from the 50276 level the most appropriate label for the
50276 level is 4s$^{2}$ $^{1}$S$_{0}$, as suggested by LBT. The observed isotope shift,
which is of about the same size as for both 3d$^9$4p-3d$^{8}$4s$^{2}$ and 
3d$^{8}$4s4p-3d$^9$4d transitions, indicates level mixing but not the size of it. 
As pointed out
by LBT an additional part of the problem may be caused by interaction with the
unknown p$^{2}$ configuration, which contains both a $^{1}$S$_{0}$ and a $%
^{3}$P$_{0}$ level. Anyhow, the fact that the 50276 level contains a substantial
amount of triplet character illustrates clearly the non-physical meaning and
sometimes excessive ambition to assign an LS label to every energy level.
However, it may be practical for book-keeping purposes to have a symbolic
name on all energy levels.

\section{Conclusion}

The $\lambda$6300 [O I] line is one of the very few spectral features from which
the oxygen abundance in the sun and cool stars (Allende Prieto et al. 2001) can be 
determined. The line is blended with an Ni I line, and a detailed study of the Ni I line 
at 6300.34 \AA\  reveals an unresolved isotope structure. At least two isotopic line 
components, separated by 20 m\AA, have to be included when calculating a line profile 
to match the $\lambda$6300 feature in high resolution stellar and solar spectra. The
log $gf$-values for the individual components should then be weighted by the relative
 abundances of the two isotopes.

We report in this paper on the wavelengths and log $gf$-values for the two major isotopic
components of the Ni I line. We have measured the radiative lifetime of the upper 
level of the Ni I line using laser techniques and combined it with branching fractions 
to get the absolute transition probabilities. The branching fractions are derived 
from calibrated line intensities in a Fourier transform spectrum, which is also used 
to disentangle the isotopic structure. The new experimental log $gf$ value differ by 0.2 dex
from the "astrophysical log $gf$ value" derived by Allende Prieto et al (2001).

An apprporiate labelling of the upper level of the Ni I line has previously been 
discussed by Litz\'{e}n et al. (1993). The additional information obtained in the 
present work, lifetime and isotopic structure, supports the label suggested in that
paper.

\end{document}